\documentclass[twocolumn]{aastex631}
\usepackage{stfloats}
\usepackage{float}
\usepackage{graphicx}
\usepackage{natbib}
\usepackage{hyperref}
\usepackage{amsmath}
\usepackage{autobreak}
\usepackage{multirow}
\usepackage{makecell}
\usepackage{color}
\usepackage{bm}
\usepackage{appendix}
\usepackage{threeparttable}
\usepackage[figuresright]{rotating}
\usepackage[mathscr]{euscript}

\received{}
\revised{}
\accepted{}
\submitjournal{}

\shorttitle{}
\shortauthors{}

\graphicspath{{./}{figures/}}

\begin{document}

\title{Internal calibration of LAMOST and Gaia DR3 GSP-Spec stellar abundances}

\correspondingauthor{Zexi Niu, Haibo Yuan}
\email{nzx@nao.cas.cn, yuanhb@bnu.edu.cn}

\author[0000-0002-3651-0681]{Zexi Niu}
\affiliation{University of Chinese Academy of Sciences, Yuquan Road, Shijingshan District, Beijing, China}
\affiliation{National Astronomical Observatories,
Chinese Academy of Sciences, 20A Datun Road, Chaoyang District, Beijing, China}

\author[0000-0003-2471-2363]{Haibo Yuan}
\affiliation{Institute for Frontiers in Astronomy and Astrophysics, Beijing Normal University, Beijing, 102206, China}
\affiliation{Department of Astronomy,
Beijing Normal University, 19th Xinjiekouwai Street, Haidian District, Beijing, China}

\author{Jifeng Liu}
\affiliation{National Astronomical Observatories,
Chinese Academy of Sciences, 20A Datun Road, Chaoyang District, Beijing, China}
\affiliation{University of Chinese Academy of Sciences, Yuquan Road, Shijingshan District, Beijing, China}
\affiliation{Institute for Frontiers in Astronomy and Astrophysics, Beijing Normal University, Beijing, 102206, China}

\begin{abstract}

{Stellar chemical abundances are crucial and fundamental in astrophysics.
However, they could suffer from substantial systematic errors according to several investigations but still lack calibrations in bulk.
By using Gaia wide binaries, we find the temperature-dependent bias between the two binary components for [Fe/H] and [$\alpha$/Fe] measurements from the LAMOST low-resolution spectra and Gaia RVS spectra.
At T$_{\rm eff}=4000$\,K, the LAMOST [Fe/H] is significantly underestimated by approximately 0.4\,dex, compared with its typical uncertainty of 0.1\,dex. 
Its [$\alpha$/Fe] is overestimated by about 0.2\,dex.
For Gaia, the underestimation of [M/H] and overestimation of [$\alpha$/Fe] becomes pronounced near 7000\,K with smaller magnitudes. 
We perform an internal calibration by minimizing the differences between binary components and provide the correction curves.
After corrections, the standard deviations of the residuals compared to the PASTEL catalog decrease from about 0.045/0.1 to 0.02/0.043 for LAMOST and Gaia, respectively. 
The chemical homogeneity of the open cluster M 44 is also improved by a factor of two. 
We stress that the underestimation of [Fe/H] could lead to an overestimation of binary fractions when selecting binary stars by the excess of luminosity. 
The method of this work could be applied to other data-sets in the future. 
Our results will benefit statistic studies that use LAMOST and Gaia samples with a wide temperature range. }
\end{abstract}

\section{Introduction} \label{sec:intro}
We have now entered a new era in which stellar parameters are being obtained on an ``industrial scale'' \citep{2019Jofr}, particularly for FGK-type stars.
{For example, RAVE \citep{rave}, SDSS/SEGUE \citep{2009Yanny}, APOGEE \citep{apogee}, LAMOST \citep{zhao2012}, and GALAH \citep{galah} spectroscopic surveys have provided spectra for more than 10 million stars.}
Recently, Gaia Data Release 3 (DR3) was published, with astrophysical parameters serving as highlights and new components \citep{2022dr3sum}. 
Astrophysical parameters for over 450 and 5 million sources were derived using BP/RP and RVS mean spectra, respectively. {Those efforts result in a dramatic increase in the data volume of stars with measurements of stellar metallicity (i.e. [Fe/H]), effective temperature (T$_{\rm eff}$), and surface gravity (log $g$). 
They are essential parameters in stellar models for calculating other properties, such as age and mass. 
With the help of enormous stellar metallicity and ages, Galactic archaeology \citep{2002Freeman} has developed rapidly in recent years and recast some long-standing questions about the formation and evolution of our Galaxy (i.e. \citealp{2011Casagrande,2018Feuillet,2019Di,2020Fragkoudi,2022Xiang}).}

{However, recent studies have revealed internal inconsistencies in the [Fe/H] measurements of individual spectroscopic surveys.
In \citet{2022Soubiran}, they examined the [Fe/H] measurements of several spectroscopic surveys by comparing them with independent determinations of [Fe/H] from the PASTEL catalog \citep{2016Soubiran}, and used FGK members in open clusters and globular clusters to access the internal consistency of [Fe/H].
The [Fe/H] residuals of LAMOST demonstrated a clear oscillation with the temperature, with the negative offsets approaching $-$0.3\,dex and $-$0.15\,dex when T$_{\rm eff}$ = 4000\,K and T$_{\rm eff}$ = 6500\,K, respectively.
This temperature-dependent bias was also observed by \citet{2019Andrews} who worked on the chemical homogeneity of wide binaries with APOGEE stars and found that when selecting binaries with $\Delta$ T$_{\rm eff}<100$\,K, the chemical abundance consistency of APOGEE  for the best-measured elements typically increases from 0.1\,dex to 0.05\,dex.}
Such systematic effects are not unexpected, as measuring the metallicities of cool stars is challenging due to the complex molecular absorption features in their optical spectra. 
{It is therefore necessary to validate and calibrate spectroscopic stellar metallicities.}

Stars belonging to wide binaries, multiple systems, and clusters are ideal tools for chemical diagnosis. 
Theoretical studies suggested that most of them were born together from the same gas cloud and at approximately the same time \citep{1997Elmegreen,2014Feng,2004Fisher,2010Kouwenhoven}. 
Therefore their initial chemical abundances should be nearly identical. 
{This has been confirmed by several studies using wide binaries \citep{2004Desidera,2020Hawkins,2021Nelson} and open clusters  \citep{2007Silva,2016Bovy} based on high-resolution and high signal-to-noise ratio spectra.}
They demonstrated that the chemical homogeneity of [Fe/H] is better than 0.05\,dex.
Taking advantage of this feature, many studies have been devoted to improving the precision of M dwarf metallicity. 
For example, \citet{2005Bonfils} determined the metallicities of 20 M dwarfs from the optical spectral of the primary and analyzed the metallicity-dependent mass-luminosity relation of the late-type star. \citet{2010Rojas} and \citet{2012Terrien} studied planet-hosting M dwarf with metallicity estimates derived from their FGK companions using near-infrared {spectra}. \citet{2020Birky} provided stellar-parameter labels for 5875 M dwarfs by training a data-driven model calibrated with metallicity from FGK primaries.
There are still rare works that could calibrate spectroscopic metallicities in bulk.

To further investigate the systematic effects of spectroscopic metallicities, we perform diagnostics of [Fe/H] and [$\alpha$/Fe] and try to establish internal calibrations for FGK dwarfs.
A large number of wide binaries that have been identified through high-quality astrometric data published by Gaia (e.g, \citealp{2020Hartman,2017pw}) makes this work feasible.
We introduce the wide binaries and spectroscopic survey used in our study in Section \ref{secdata}. 
The diagnostic and internal calibrations are demonstrated in Section \ref{secsys}. We discuss the assumption of homogeneous metallicity in section \ref{secch}. We validate results in Section \ref{secpastel} and \ref{secvad}. Finally, methods and results are summarized in Section \ref{summary}.

\section{Data properties}\label{secdata}

\subsection{Wide binary}

We use the wide binary catalog published by \citet{2021wide}, where over 1 million binaries within 1 kpc and with projected separations from a few au to 1 pc were identified using Gaia EDR3 dataset \citep{gaiaedr3} through the common proper motion technique. 
{Contamination from} background pairs, clusters, and triples have been carefully discarded. 
They evaluated the rate of chance alignments (\textit{R}) as a function of observables. 
The larger the \textit{R}, the higher the rate of contamination.
{By applying \textit{R} $< 0.1$ to select pairs with 90 percent bound probability, 0.5 million out of 1.8 million samples are removed.
They labelled stars as WDs or MS (including giants) according to the Gaia color-absolute magnitude diagram (CMD).
Here we also limit samples to MS+MS type, resulting in 877,416 pairs.}

\subsection{LAMOST LRS}

The low-resolution spectroscopic survey (LRS) of LAMOST began in 2011 and has accumulated spectra for more than 10 million stars until DR8 \citep{2022Yan}.
The spectra range from 370 to 900 nm with $R=1,800$. 
A number of stellar parameter pipelines from different research groups have been developed. 
Here we focus on two versions. 
One is from the official LAMOST stellar parameter pipeline (LASP) \citep{2011Wu}. 
It determines the stellar parameters (T$_{\rm eff}$, log $g$, [Fe/H]) by minimizing $\chi^2$ between observed spectra and model spectra from the ELODIE library and achieves the precision of about 110\,K, 0.2\,dex, and 0.1\,dex, respectively \citep{2015luo}.
Here we adopt the LASP DR8 which contains stellar parameters of over 4.7 million stars.
Another one is from the data-driven Payne (DD-Payne) approach for LAMOST DR5 \citep{xiangddp}. 
The machine-learning techniques are applied to train the stellar labels of APOGEE DR14 and GALAH DR2 into LAMOST labels. 
The internal precision of T$_{\rm eff}$, log $g$, and [Fe/H] are typically 60\,K, 0.15\,dex, and 0.05\,dex for S/N $>$ 20.
In addition, [$\alpha$/Fe] is provided as a weighted mean of [Mg/Fe], [Si/Fe], [Ca/Fe], and [Ti/Fe] with a precision of about 0.03\,dex.

\begin{figure}[htbp] 
    \centering 
    \includegraphics[width=3.5in]{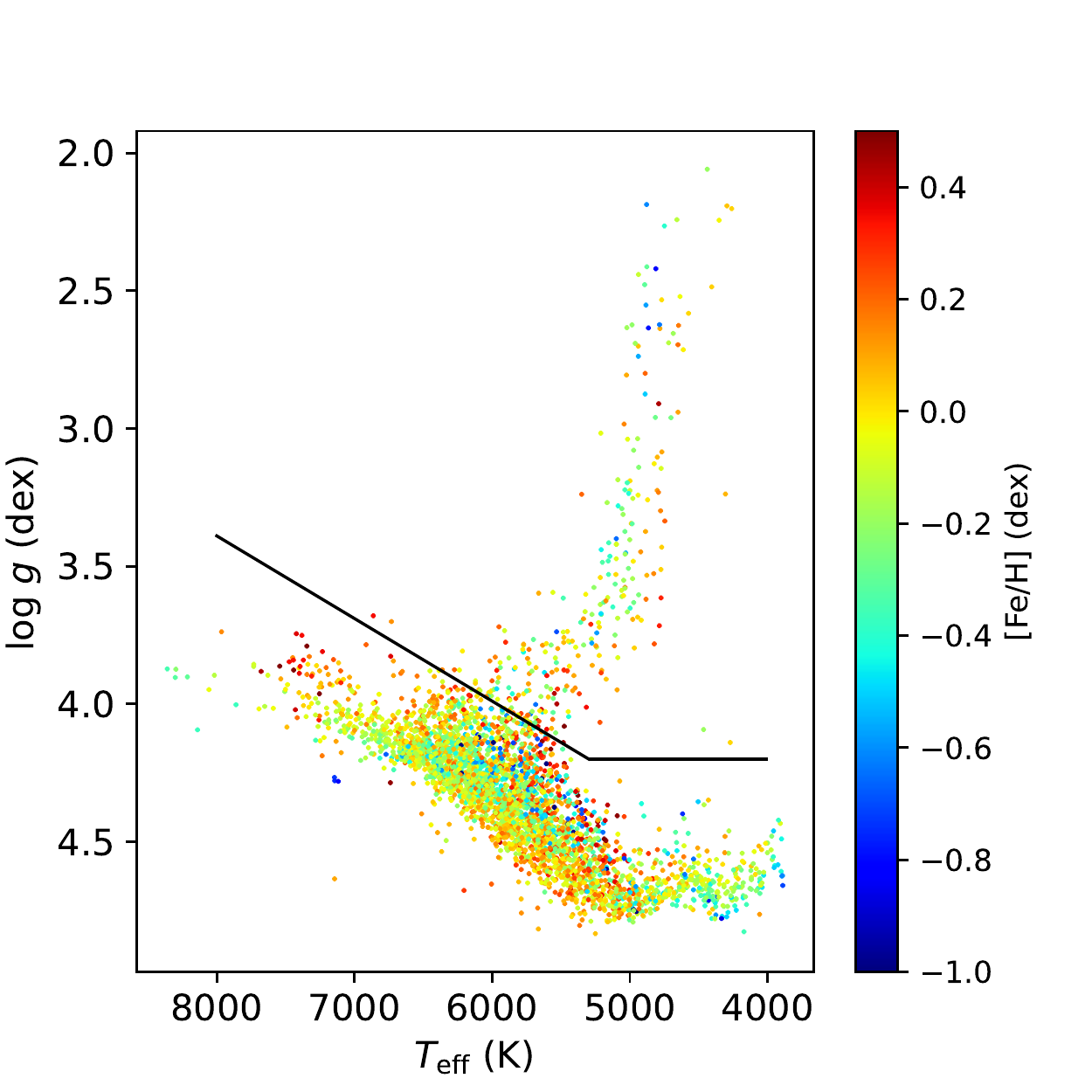}
    \caption{Kiel diagram with parameters from the LAMOST LASP of member stars of wide binaries. Stars above the black lines were removed.
    \label{kiel}} 
\end{figure}

\begin{figure*}[htbp] 
    \centering 
    \includegraphics[width=7.5in]{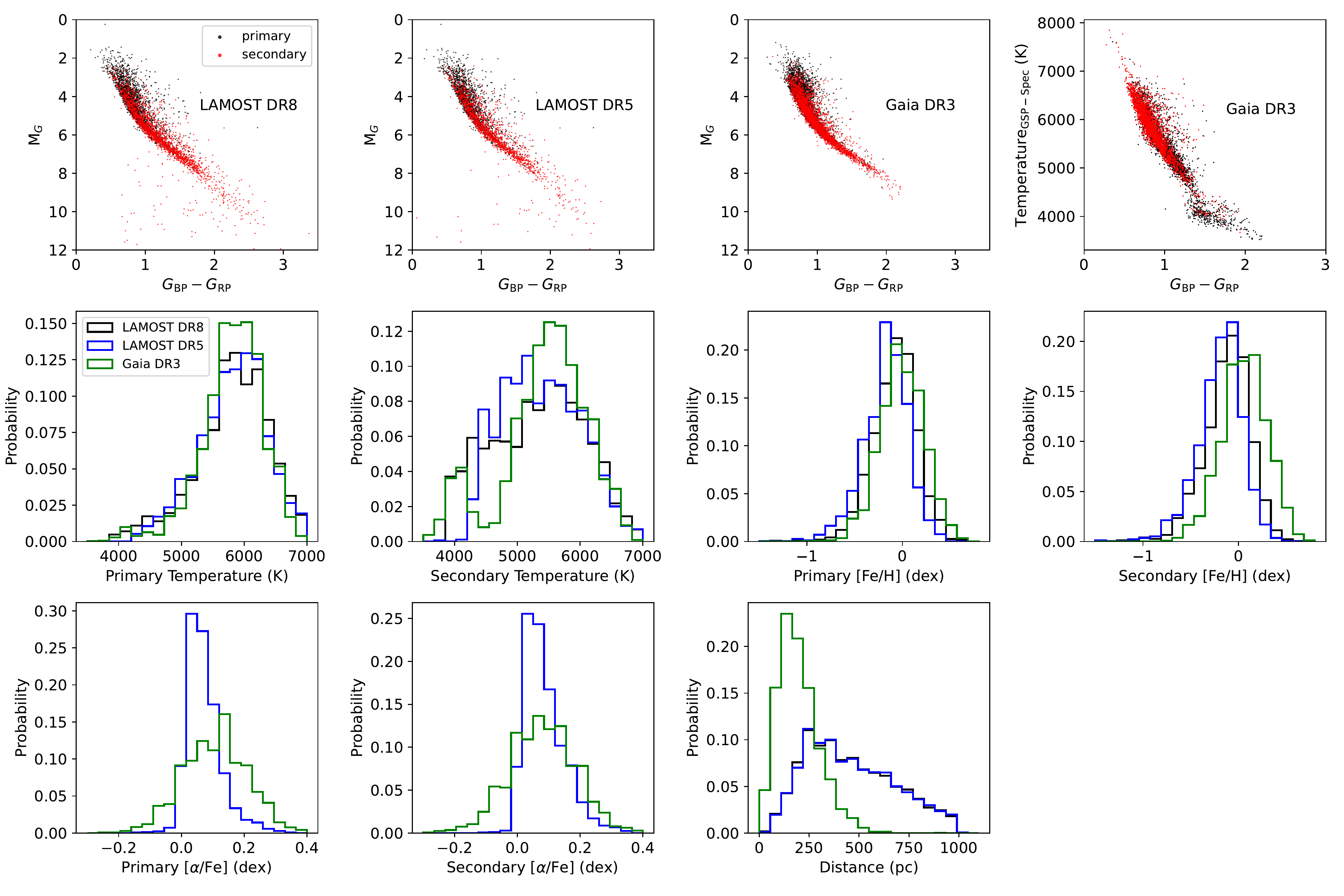}
    \caption{{Basic properties of the wide binaries used in this work: color absolute-magnitude diagrams (defining M$_G = G$ + 5log($\varpi$/1000)), temperature, [Fe/H], [$\alpha$/Fe], distance. The two LAMOST samples behave quite similarly. Compared with the LAMOST samples, the Gaia sample has a smaller distance due to the limitation of $G < 16$ from the Gaia. We noticed that the temperature determination from GSP-Spec has a systematic offset nearing 4000\,K. We kept this part of the sample and left this problem to future work.}
    \label{cmd}} 

\end{figure*}

We cross-matched LASP DR8 and DD-Payne DR5 with the selected wide binary sample and required S/N $>$ 20. 
For stars with multiple observations, values with the highest S/N were chosen. 
{Since \citet{2021wide} defined MS according to the CMD, we removed giant stars and selected dwarf stars from the T$_{\rm eff}$--Log $g$ (Kiel) diagram as shown in Figure \ref{kiel}.}
The final samples included 2,296 pairs for LASP and 1,698 pairs for DD-Payne in which both components satisfy the requirements. 
{Their distributions of temperature, [Fe/H], [$\alpha$/Fe], distance, and CMDs are displayed in Figure \ref{cmd}. 
The LASP and DD-Payne samples have quite similar properties except that the DD-Payne sample contains fewer secondaries with $T_{\rm eff}< 4200$\,K and have systematically smaller [Fe/H] which may be inherited from its training sets.}
Most of their temperatures range from 4000\,K to 7000\,K.
Their [Fe/H] and [$\alpha$/Fe] cover $-1$ $\sim$ 0.5\,dex and 0 $\sim$ 0.2\,dex, respectively. 

\subsection{Gaia GSP-Spec}

In Gaia DR3, atmospheric parameters are impressively introduced as part of new data products \citep{2022dr3sum}. 
Gaia GSP-Spec derives atmospheric parameters for over 5 million stars based on the RVS spectra which cover 845 - 872 nm with $R=11,500$. 
It operates on all sources with spectral S/N$>$ 20, which approximately corresponds to stars brighter than 14 magnitudes.  
Spectra are fitted by the synthetic spectra based on MARCS models using Matisse-Gauguin and ANN workflows separately to estimate T$_{\rm eff}$, log $g$, [M/H], and [$\alpha$/Fe]. 
{Notice that the overall metallicity ([M/H]) derived from Gaia GSP-Spec resembles the [Fe/H] obtained from LAMOST. For the sake of convenience, in the following text, we use the [Fe/H] to refer to [M/H] when they appear together.}  
We focused on the results that come from the Matisse-Gauguin algorithms and are presented in \texttt{astrophysical$\_$parameters} table.
More details on the parametrization can be seen in \citet{2022gspspec} and \citet{2022Fouesneau}.
We {selected} common objects of the GSP-Spec and the selected wide binaries and further limited data quality by setting \texttt{flags$\_$gspspec[0:8] $\leq$ 1} \citep{2022gspspec}. 
We stress that dwarfs make up a small fraction when the data quality is carefully restricted because of their low intrinsic brightness.
After selecting dwarfs, we got 3,427 pairs. 
{As shown in Figure \ref{cmd}, their spectral type is similar to that of the LAMOST.
They cover wider ranges in metallicities ranging from $-$1 to 0.75\,dex for [M/H] and $-$0.35 to 0.45\,dex for [$\alpha$/Fe].
Because Gaia only executes GSP-Spec for stars with $G < 16$ mag, our samples are mostly around 200 pc. We noticed that the temperature distribution of Gaia secondaries exhibits two peaks, where the low-temperature peak around 4000K may be attributed to a systematic offset. There is a marked discontinuity appears around $G_{\rm BP} - G_{\rm RP} \sim 1.4$ and $T_{\rm GSP-Spec} \sim 4000$\,K in the top right subfigure of Figure \ref{cmd}. Considering the number of stars nearing 4000\,K is small, we kept them and did not do any extra processing.}

\section{Temperature-dependent bias}\label{secsys}

\subsection{Performance}

\begin{figure*}[htbp] 
    \centering 
    \includegraphics[width=6in]{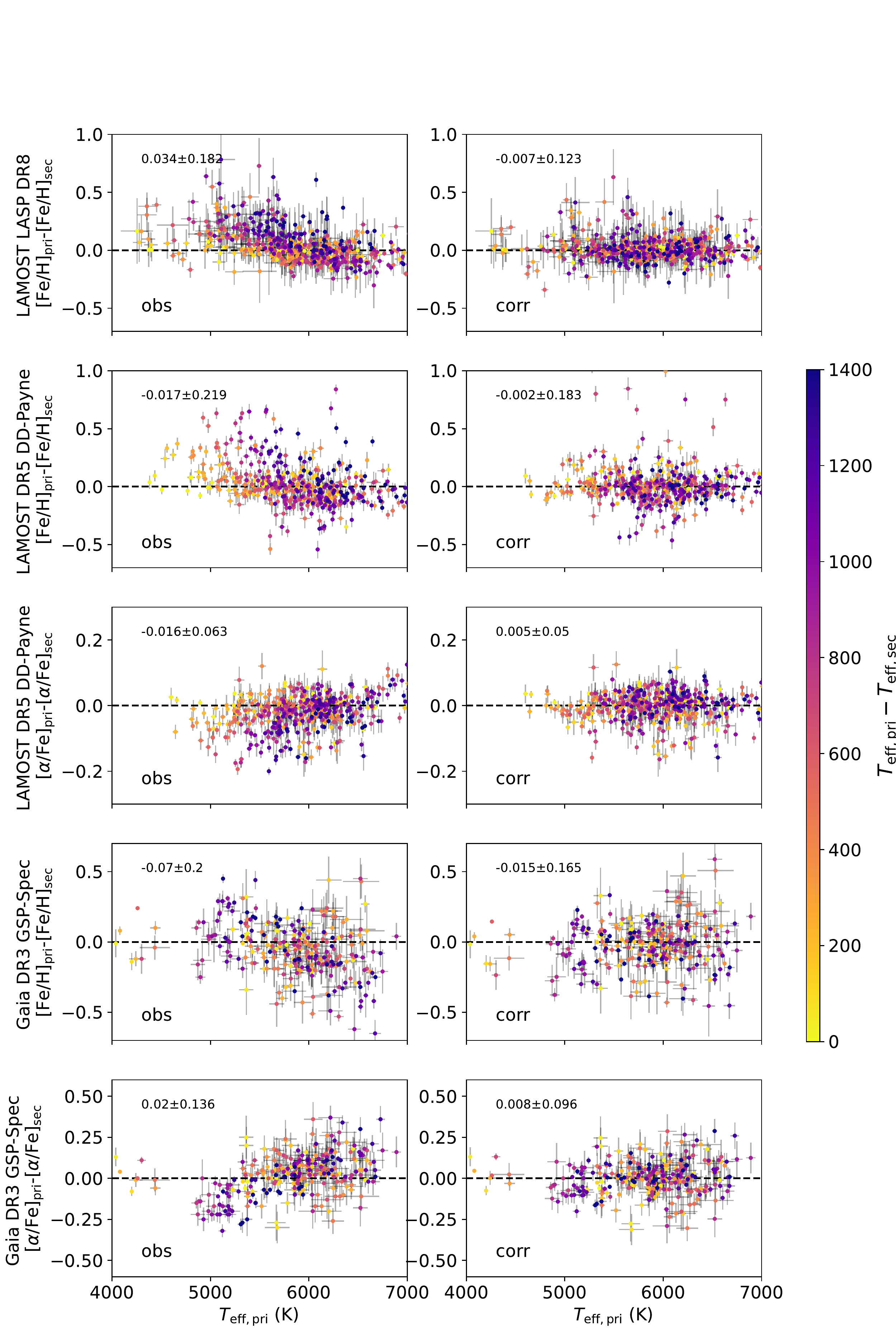}
    \caption{Differences between the chemical abundances of wide binary components as a function of the temperature of the primary. 
    Colors mark the differences in temperatures. Error bars are over-plotted.
    {To emphasize the effect of the correction, the average residual and standard deviation for binaries with  $\Delta$T$_{\rm eff} > 600 \ $K are texted in each panel.}
    The left column is before corrections, the right column is after corrections.
    Part of ``twin" binaries, namely $\Delta$T$_{\rm eff} < 200 \ $K, are not drawn for clarity. 
    \label{dfeh}} 
\end{figure*}

Here we use wide binary stars to show the {temperature-dependent} bias in stellar abundances.
We defined the hotter component as the primary and calculated differences in temperatures and metallicities of each binary system.
{The left panels of Figure \ref{dfeh}}  demonstrate $\Delta$[Fe/H] (or $\Delta$[M/H]) and $\Delta[\alpha$/Fe] as a function of temperature of the primary. 
As we can see from all panels, yellow dots that represent binaries containing two components with similar temperatures are distributed around zero at all $T_{\rm eff,pri}$.
However, purple dots that represent binaries with notable temperature differences show a temperature-dependent distribution. 
For [Fe/H] of LAMOST DR5 and DR8, the purple dots show similar tendencies except for a sharper transition in LAMOST DR5. 
Overall, the significant offset occurs approximately below 6000\,K, the lower the $T_{\rm pri}$ of the purple dots, the larger the deviation in $\Delta$[Fe/H] to zero. 
This variation suggests that the [Fe/H] of cold stars is systematically lower than that of hot stars when the temperature is below 6000\,K.
We stress that the highest offset appears around 5500\,K does not mean that stars colder than 5500\,K are less biased.
A purple dot with $T_{\rm pri} \sim$ 5500\,K corresponds to that the temperature of the secondary is near 4000\,K.
Below that, pairs are just dominated by binaries with close temperatures due to the selection limitation.
Besides, for $T_{\rm pri}$ higher than 6000\,K, the temperature-dependent bias is less obvious, which indicates that the [Fe/H] of hot stars is slightly underestimated compared to that of stars near 6000\,K.

The [$\alpha$/Fe] of LAMOST DR5 resembles the [Fe/H] but with an opposite slope. 
That is, the [$\alpha$/Fe] is overestimated at most temperatures. Its magnitude first decreases and then increases from 4000\,K to 7000\,K, the transition occurs around 6000\,K too. 

For Gaia DR3,  the smallest $\Delta$[M/H] is around 6300\,K and gradually increases towards the hot and cold end,  which seems like the shape of [Fe/H] for LAMOST DR5. 
The [$\alpha$/Fe] of Gaia DR3 looks similar to that of LAMOST DR5 but with a much larger dispersion.
It's worth mentioning that the large dispersion of metallicity of Gaia DR3 is clearly exhibited compared with the LAMOST. 
The typical dispersion of yellow dots is 0.1\,dex for LAMOST, whereas 0.2\,dex for Gaia DR3, which reflects the larger typical uncertainty of Gaia DR3. 
We suppose that the low S/N on average \footnote{The S/N of Gaia RVS spectrum can be found as \texttt{rv$\_$expected$\_$sig$\_$to$\_$noise} and \texttt{rvs$\_$spec$\_$sig$\_$to$\_$noise}. The second is provided only for the stars having the \texttt{rvs$\_$mean$\_$spectrum} published.} and narrow wavelength coverage could account for it.

\begin{figure}[htbp] 
    \centering 
    \includegraphics[width=3.5in]{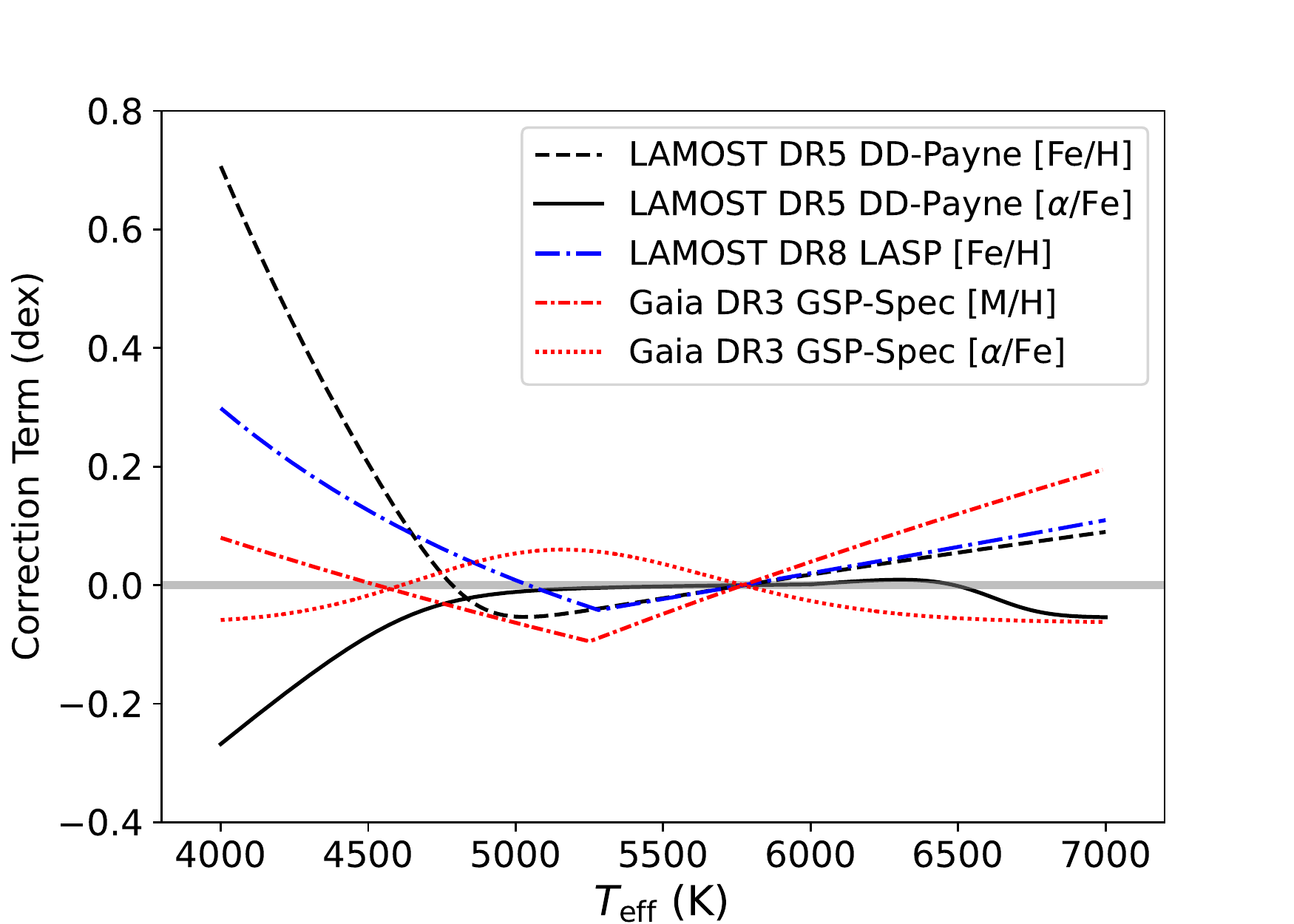}
    \caption{Temperature-dependent correction terms for the LAMOST and Gaia DR3 GSP-Spec stellar abundances. {A gray solid line at 0.0 is over-plotted for reference.}
    \label{cc}} 
\end{figure}

\subsection{Internal calibration}\label{secmethod}

Instead of external calibrations that depend on literature data, we perform internal calibrations of stellar abundances.  
In general, the calibration curves are obtained by minimizing the abundance differences between the two components.
The cost function can be written as:
{$([{\rm A/B}]_{\rm pri} + f(x))^2 - ([{\rm A/B}]_{\rm sec} + f(x))^2$}. Here $x$ = T$_{\rm eff}$ and $f(x)$ represents the temperature-dependent corrections.
The forms of $f(x)$ come from the analysis in the previous section and are described in the following text for each survey.
A smoothly broken power law was applied to the [Fe/H] of LAMOST DR8 and [$\alpha$/Fe] of Gaia DR3 as shown in Equation \ref{eqsbpl}, where $x_b$ is the breaking point; $\Delta$ is the smoothness parameter; $\alpha_1$ and $\alpha_2$ is the power law index for $x \gg x_b$ and $ x \ll x_b$, respectively; {$C$ is a constant term assuming the measurement of the metallicity of solar-like stars is unbiased}.
Given that the shapes of [Fe/H] of LAMOST DR5 and [M/H] of Gaia DR3 resemble the LAMOST DR8 but with an abrupt transition, a simple broken power law was used as written in Equation \ref{eqbpl}.
When calibrating the [$\alpha$/Fe] of LAMOST DR5, we found that one broken power law cannot work for both high and low temperatures. 
So we first corrected the two sides separately and then connected them by making the correction values equal at 6000\,K. 

\begin{equation}\label{eqsbpl}
    f(x)=A\left(\frac{x}{x_b}\right)^{-\alpha_1}\left\{\frac{1}{2}\left[1+\left(\frac{x}{x_b}\right)^{1/ \Delta}\right]\right\}^{(\alpha_1-\alpha_2)\Delta}+C
\end{equation}

\begin{equation}\label{eqbpl}
    f(x)=
    \left\{
             \begin{array}{lr}
             A(x/x_b)^{-\alpha_1}+C : x < x_b &  \\
             A(x/x_b)^{-\alpha_2}+C : x > x_b 
             \end{array}
\right.
\end{equation}

We have noticed that about two-thirds of the pairs have very close temperatures (e.g. $\Delta$ T$_{\rm eff} < 200$\,K). Details about this ``twin" phenomenon can be seen in several works (e.g. \citealp{2019twin}).
In order to avoid the twin binaries dominating the fitting, we generated the fitting samples randomly so that their temperature differences are flatly distributed.   
In each fitting, a 64-walker and 10,000-step Markov chain Monte Carlo (MCMC) method was taken with the Python package \texttt{emcee} \citep{emcee}. 
We repeated the above procedure 50 times to reduce the random error and set the medium values as the coefficients.
We set $f(5770) = 0$ to get the $C$.
All coefficients are summarized in Table \ref{coeff} and plotted in Figure \ref{cc}.

The fitted curves are as expected. 
The [Fe/H] of LAMOST is significantly underestimated for stars colder than 4500\,K where the systematic errors clearly exceed the typical uncertainties.
Meanwhile, a modest underestimation also exists for stars hotter than 6500\,K. 
While for the [M/H] of Gaia DR3, a modest underestimation occurs on the cold side and an obvious underestimation happens on the hot side.
Unlike [Fe/H],  [$\alpha$/Fe] are overestimated in most cases. The magnitude of Gaia DR3 is smaller than that of LAMOST DR5, which reaches 0.2\,dex at the low-temperature end.

We applied the calibration curves to our wide binary samples and plotted the abundance differences with the corrected ones in the right column of Figure \ref{dfeh}. 
The temperature-dependents bias is eliminated successfully.
We will further validate the calibrations in the next section. 

Notice that the corrections are based on dwarf stars of 4000\,K $\sim$ 7000\,K, extrapolations beyond should be used with caution.  

\begin{table*} [htbp]
    \centering
    \caption{Fitting coefficients of the internal calibration of the chemical abundances. \label{coeff}}
    \begin{tabular}{l|c|c|c|c|c|c}
     & $A$ & $x_b$ & $\alpha_1$ & $\alpha_2$ & $\Delta$  & $C$ \\
     \hline

LAMOST DR5 [Fe/H] & 2.171 & 4860.9 & 1.610 & $-$0.207 & 0.017 &  $-$2.202   \\
\hline

LAMOST DR8 [Fe/H] & 0.358 & 5281.4 & 2.404 & $-$1.254 &  --  & $-$0.400\\
   \hline

Gaia DR3 [M/H] & 16.033 & 5250.9 & 0.040 & $-$0.062 & -- & $-$16.127 \\
\hline
LAMOST DR5 [$\alpha$/Fe]$_{< 6000 {\rm K}}$ & 3.792 &  4669.5  & $-$0.454 & $-$0.010 & 0.032 &  $-$3.837  \\
   \hline
LAMOST DR5 [$\alpha$/Fe]$_{> 6000 {\rm K}}$ & 0.022 & 6684.0  &  $-$4.035 & 107.143 & 0.018  & $-$0.055 \\
\hline
Gaia DR3 [$\alpha$/Fe] & 0.117 & 5319.6 & $-$76.871 & 83.418 & 0.373 & $-$0.063 \\

    \end{tabular}
\end{table*}

\section{Discussion}\label{secdis}

\subsection{Chemical Homogeneity of wide binaries}\label{secch}

As mentioned previously, this work is based on the assumption that the chemical abundances of two binary components are identical. 
However, in some cases, this assumption might be challenged by binarity, atomic diffusion, and exoplanet. 
We discuss these factors below.

By analyzing a volume-limited sample of 4847 stars within 67 pc, \citet{2014Tokovinin} suggested that the hierarchical multiplicity of F and G Dwarfs is $13\pm1\%$.
\citet{2022Hartman} further investigated the overluminous stars from the wide binary catalog of \citet{2020Hartman} and deduced that the lower limit of the higher-order multiplicity of K$+$K wide binaries is $40\pm1.6\%$. 
The presence of an unresolved faint secondary can introduce systematic bias for stellar parameters when fitting spectra with a single-star model.
Normally, the temperature and metallicity will be underestimated, where the scale varies with the temperature of the companion star and the mass ratio.
This has been simulated by \citet{bi_specfit}. 
They modeled spectra similar to APOGEE, GALAH, and LAMOST surveys and found the impacts are negligible for the stellar parameters derived from the low/middle-resolution spectra like LAMOST with $R = 1,800$ and S/N$ = 30$. 
The effect on GALAH-like spectra with $R = 28,000$ and S/N$ = 100$ is marginally larger than the LAMOST, probably because of its narrow wavelength coverage, $\sim$ 100 nm in total. 
Although Gaia RVS spectra have $R = 11,500$, the typical uncertainties of Gaia GSP-Spec are larger than those of LAMOST, the resulting errors for Gaia GSP-Spec should be inconsequential as well. 
Even considering that the binarity would lower the observed metallicity and temperature, the binary fraction is anti-correlated with the stellar temperature \citep{2019Moe,2021Niu}, which would affect the cold stars slightly and cannot explain the result in Section \ref{secsys}.
We suppose the binarity barely affects our results.

In addition to the systematic bias caused in the data analysis, the current surface compositions of the FGK dwarfs can differ from the initial ones due to several physical processes such as atomic diffusion and formation or decomposition of exoplanets. 
Atomic diffusion lowers the surface abundance of heavy elements, which will reduce the value of [Fe/H]. 
Its magnitude is tightly correlated with stellar mass and age, as shown in \citet{2017Dotter}, 
In principle, lower-mass stars have deeper convection zones therefore weaker diffusion effects.
For stars with $M_{\rm init} =1 M_{\rm \odot}$ and [Fe/H]$_{\rm init} = 0$, the diffusion effect is expected to be less than 0.08\,dex, which is comparable to binary effect but negligible for the [Fe/H] derived from the low/middle-resolution spectra.
As for [X/Fe] such as [$\alpha$/Fe], since the elements are affected by roughly the same level \citep{2016Choi,2019Souto}, the diffusion effects have been virtually eliminated \citep{2022Souto}.
Another possible channel that could cause different metallicities is related to exoplanets by either accreting or taking away materials.
However, whether exoplanets significantly change the surface abundance of the host star is still controversial (e.g. \citealp{2021Liu}). Despite the huge diversity, the occurrence rate of FGK-type stars having an Earth-like planet is less than 35 $\%$ \citep{2015Winn}. 
The rate declines substantially towards larger planet radius \citep{2012Howard}.
The resulting magnitude tightly varies with the type, number, and size of the planet, at least the scale is far less than the bias shown in Section \ref{secsys}. 
Considering the bias is tightly distributed, it prefers systematic effects instead of effects from exoplanets that should have a larger dispersion.

\subsection{PASTEL catalog}\label{secpastel}

\begin{figure}[htbp] 
    \centering 
    \includegraphics[width=3.7in]{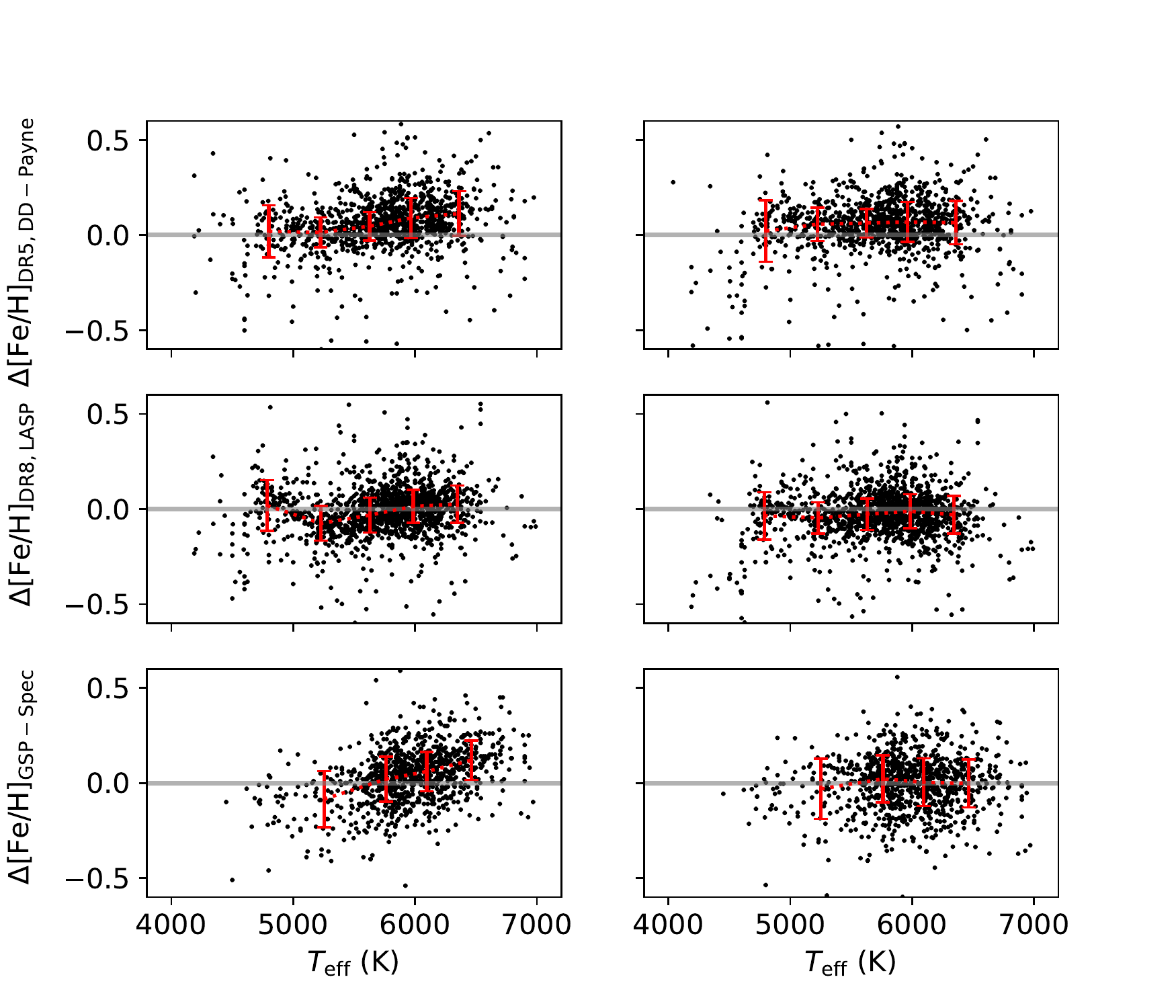}
    \caption{Differences of [Fe/H] between the PASTEL catalog \citep{2016Soubiran} and surveys investigated in this work. 
    Abundances before and after calibration are plotted in the left and right columns, respectively.
    Red dots and bars represent median values and standard deviations. 
    $\Delta$[Fe/H]=0 is plotted in gray lines. 
    \label{pastel}} 
\end{figure}

We validate the calibrations by comparing them with the well-calibrated stellar parameter catalogs, such as the PASTEL catalog \citep{2022Soubiran}. 
Their [Fe/H] measurements came from high-resolution spectra where metal lines are resolved. 
In such cases, the degeneracy between temperature and metallicity is weak.
But for low/middle-resolution spectra, iron lines are blended, astrophysical parameters are derived through the global spectral fitting. We only focus on [Fe/H] because [$\alpha$/Fe] is not available in the PASTEL catalog.
Samples in Figure \ref{pastel} are generated following the same procedure in Section \ref{secdata}.
{Red dots and bars trace how the [Fe/H] differences vary with temperature. 
A gray solid line at $\Delta$[Fe/H]=0.0 is over-plotted for reference.
A pronounced temperature-dependent bias of $\Delta$[Fe/H] is seen before the correction in the left column. 
The standard deviations of their red dots are 0.047, 0.045, and 0.10 from the top to the bottom panel.
While after the correction, the temperature-dependent tendency barely exists, where the standard deviations of red dots decrease to 0.027, 0.018, and 0.043, showing a good consistency between the corrected [Fe/H] and the PASTEL measurements. 
Notice that we did not try to obtain absolute corrections therefore the average residuals could deviate from zero.}

\subsection{Open Cluster}\label{secvad}

\begin{figure}[htbp] 
    \centering 
    \includegraphics[width=3.2in]{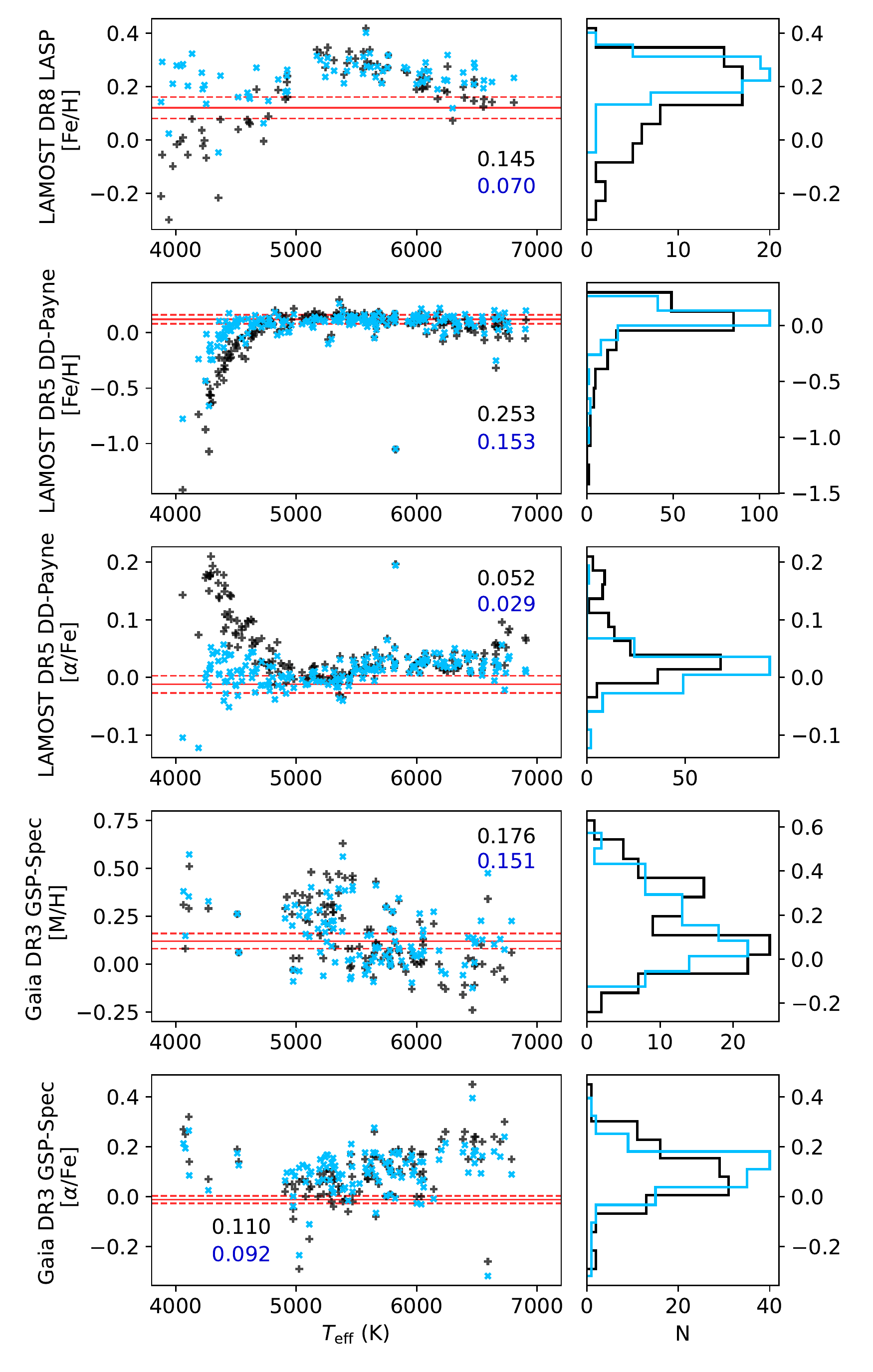}
    \caption{Left: Chemical abundances as a function of stellar temperatures for open cluster M 44 member stars. 
    The original and revised abundances are colored in black and blue, respectively. Their standard deviations are texted in black and blue, respectively.
    Red lines are the literature values of solar-type dwarfs from \citet{2013Boesgaard}.
    Right: Histogram distributions of the chemical abundances.
    \label{oc}} 
\end{figure}

Stellar clusters are commonly used to validate the measurements of stellar chemical abundances, parallaxes, and ages due to their simple properties of being co-natal and coeval, {namely the same birthplaces and ages.}
Therefore, we selected dwarf stars of M 44 in these three data-sets, by cross-matching the catalog from \citet{cg2020}. 
We plotted their metallicities as a function of temperatures in Figure \ref{oc}. 
Black and blue scatters indicate the metallicities before and after the internal calibrations, respectively.
{Their distributions are shown in the right column and standard deviations are text in the left column using the same color.}
The top three rows are for the LAMOST DR8 LASP and DR5 DD-Payne. 
Before the calibration, some member stars have unreasonably low [Fe/H] and high [$\alpha$/Fe] values, especially at the low-temperature end. 
After the calibration, the temperature-dependent bias is greatly eliminated except for several stars near 4000\,K.
The standard deviations of the corrected abundances are reduced by a factor of two.
The bottom two rows are for the Gaia DR3 GSP-Spec. The calibration reduces the standard deviations by about 15 percent.
The abundances of M 44 metallicities have been measured in many works. 
{Here we chose the one from \citet{2013Boesgaard}, where [Fe/H] = $0.12\pm0.04$ and [$\alpha$/Fe] = $-0.012\pm0.015$ are deduced from the $R = 45,000$ spectra of solar-type dwarfs, and plot their measurements and uncertainties in red lines.
For some panels, like LAMOST DR8 LASP, systematic offsets exist between the literature values and our results which is tied to the choice of the zero point. 
We leave the problem of absolute correction to future work.}

\begin{figure*}[htbp] 
    \centering 
    \includegraphics[width=7in]{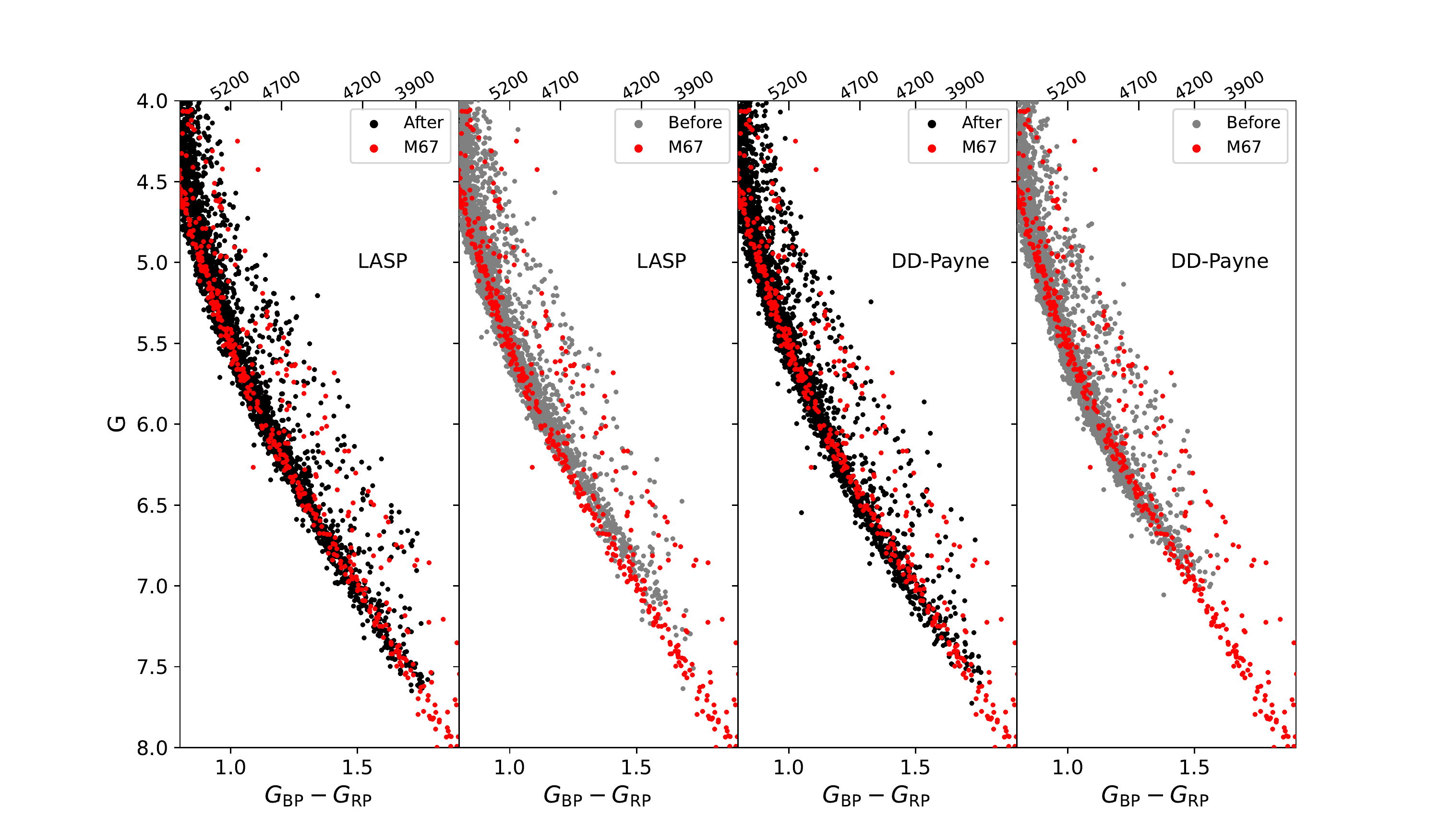}
    \caption{Comparison of the HRDs of open cluster M 67 and stars with the same [Fe/H] before and after corrections. 
    Red points are member stars of M 67 at 0.88 kpc according to \citet{cg2020}. 
   The black points match the locus of M 67 better than the gray points.
   The deviation of the gray points toward lower magnitude at T$_{\rm eff}< 4700$\,K indicates that they have larger [Fe/H] than that of red points. 
    The SFD dust map and extinction coefficients from \citet{2023Zhang} are applied.
    \label{m67hrd}} 
\end{figure*}

{Member stars of the stellar cluster also provide the high-quality Hertzsprung-Russell diagram (HRD). 
In order to address the effect of our corrections on the HRD, we plotted Figure \ref{m67hrd} with Gaia DR3 photometry. 
All points are required to satisfy $G\_flux\_over\_error > 50$, $BP/RP\_flux\_over\_error > 50$, $parallax\_over\_error > 10$, and galactic latitude $|b| > 20^{\circ}$. Black and gray points also satisfy E(B$-$V)$_{\rm SFD} < 0.02$.
Their absolute magnitudes and colors are dereddened with the SFD dust map and reddening coefficients as provided in \citet{2023Zhang}. 
Red points are member stars of M 67 that can represent the locus of stars having the same metallicity as M 67 on the HRD.
Gray and black points are selected using $|$[Fe/H]$-$[Fe/H]$_{\rm M67}| < $ 0.01 and $|$[Fe/H]$_{\rm corr}-$[Fe/H]$_{\rm M67}| < $ 0.01, respectively.
[Fe/H]$_{\rm M67}$ is adapted from Figure 11 of \citet{2022Soubiran}. 
We also considered the different zero points between LASP/DD-Payne and PASTEL. 
The corresponding temperatures are labelled at the top of all panels as well.
It is obvious that gray points gradually deviate from the red points towards lower magnitude with decreasing temperatures. 
As shown in Figure B1 in \citet{2021Niu} as well as Figure 4 in \citet{2022Hartman}, single stars place on the HRD in the order of metallicity tracks with metal-rich stars being brighter.
The deviation indicates that gray points, especially those below 4700\,K, have larger [Fe/H] in fact. 
While after the correction, black points match the red points until the cold end of the temperature. 
It reminds us that some normal single stars with T$_{\rm eff} < 4700$\,K could be mistaken as overluminous objects like binary stars for their locations above the main sequence without the corrections.
Therefore binary fractions could be overestimated when selecting binaries through the HRD.}

\section{Summary}\label{summary}

We use wide binaries to investigate the systematic errors of stellar chemical abundances in the LAMOST and Gaia databases.
We start with the Gaia EDR3 MS/MS binaries compiled by \citet{2021wide}. 
For LAMOST, the examinations are done for stellar parameters determined with the LASP of DR8 and DD-Payne of DR5.
For Gaia DR3, we test the atmospheric parameters based on the RVS spectra. 
As predicted theoretically, the chemical abundances of binary components should be homogeneous. 
However, we find inconsistencies in the three catalogs that greatly exceed their typical uncertainties, especially for pairs with large differences in temperature. 
Also, the discrepancies systematically vary with the temperature of the primary. 
After ruling out the possibilities of binarity, atomic diffusion, and exoplanet, we suggest that it belongs to the systematic error of spectroscopic metallicity.

We calibrate the bias internally by fitting a correction term as a function of the temperature to minimize the differences between the two components, in which no external information is included. 
The correction curves are listed in Table \ref{coeff} and plotted in Figure \ref{cc}.
{For FGK type dwarf stars, the underestimation of LAMOST [Fe/H] can reach 0.4\,dex at 4000\,K and 0.1\,dex at 7000\,K; LAMOST [$\alpha$/Fe] values are overestimated by about 0.2\,dex at 4000\,K.}
While for the Gaia DR3 GSP-Spec, the correction term is typically smaller than that of LAMOST, approximately 0.3\,dex for [M/H] and 0.15\,dex for [$\alpha$/Fe] at maximum. 
We stress that systematic errors have a more prominent influence on the data with small random errors.
{The corrections are validated by the PASTEL catalog as shown in Figure \ref{pastel}. After corrections, the standard deviation of [Fe/H]$_{\rm LASP/DD-Payne/GSP-Spec}$-[Fe/H]$_{\rm PASTEL}$ are reduced to 0.018, 0.027, and 0.043, which are nearly half of what they were before.
By taking M 44 as an example, we demonstrate the improvements in the chemical homogeneity of OC in Figure \ref{oc}.  
Similarly, the standard deviations of the corrected abundances are generally reduced by a factor of two. 
The impacts on the HRD are also illustrated in Figure \ref{m67hrd}.
We stress that when figuring out the overluminous objects like binary systems from the HRD, the temperature-dependent bias of the metallicity should be carefully corrected.}

To our knowledge, there were no internal calibrations concentrating on the systematic errors of LAMOST abundances yet. 
This work demonstrates the possibility of validating and calibrating stellar abundances in bulk using wide binaries, which could be applied to other data-sets in the future.  
Our results will benefit statistic studies that use enormous LAMOST and Gaia stars with a wide temperature range. 

\nolinenumbers 

\begin{acknowledgments}
We acknowledge the referee for his/her valuable comments and suggestions that improved the  quality of the paper significantly. 
We acknowledge the helpful discussions with Prof. Chao Liu and Shuai Liu. This work was supported by the National Natural Science Foundation of China (NSFC) under grant numbers
11988101, 11933004, 12222301, National Key Research and Development Program of China (NKRDPC) under grant
numbers 2019YFA0405504 and 2019YFA0405503, and
Strategic Priority Program of the Chinese Academy
of Sciences under grant number XDB41000000.
We also acknowledge the science research grants from the China Manned Space Project with NO. CMS-CSST-2021-A08 and CMS-CSST-2021-A09.

This work has made use of data products from the Guoshoujing Telescope (the Large Sky Area Multi-Object Fiber Spectroscopic Telescope, LAMOST). LAMOST is a National Major Scientific Project built by the Chinese Academy of Sciences. Funding for the project has been provided by the National Development and Reform Commission. LAMOST is operated and managed by the National Astronomical Observatories, Chinese Academy of Sciences. 

This work has made use of data from the European Space Agency (ESA) mission Gaia (https://www.cosmos.esa.int/gaia), processed by the Gaia Data Processing and Analysis Consortium (DPAC, https://www.cosmos.esa.int/web/gaia/dpac/consortium). Funding for the DPAC has been provided by national institutions, in particular the institutions participating in the Gaia Multilateral Agreement.

\end{acknowledgments}

\bibliography{sample631}
\bibliographystyle{aasjournal}

\end{document}